\documentclass{appolb}
\usepackage{epsfig}

\begin{document}
\title{Dynamics of the Warsaw Stock Exchange index as analysed by the
nonhomogeneous fractional relaxation equation\thanks{Presented at the Second
Polish Symposium on Econo- and Sociophysics. Poland, Krak\'ow 21-22 April 
2006, internet address: http://www.ftj.agh.edu.pl/fens2006/}
}
\author{Marzena Koz{\l}owska and Ryszard Kutner\thanks{E-mail: erka@fuw.edu.pl}
\address{Division of Physics Education, Institute of Experimental 
Physics \\  Department of Physics, Warsaw University \\ Smyczkowa Str. 5/7, PL-02678 Warsaw, Poland}
}
\maketitle
\begin{abstract}
We analyse the dynamics of the Warsaw Stock Exchange index WIG at a daily time horizon before and after
its well defined local maxima of the {\it cusp}-like shape decorated with oscillations. The rising and
falling paths of the index peaks can be described by the Mittag-Leffler function superposed with 
various types of oscillations. The latter is a solution of our model of index dynamics defined by 
the nonhomogeneous fractional relaxation equation. This solution is a generalised analog of an exactly solvable 
model of viscoelastic materials. We found that the Warsaw Stock Exchange can be considered as an 
intermediate system lying between two complex ones, defined by short and long-time limits of the 
Mittag-Leffler function; these limits are given by the Kohlraush-Williams-Watts law for the initial times, 
and the power-law or the Nutting law for asymptotic time. Hence follows the corresponding short- and 
long-time power-law behaviour (different "universality classes") of the time-derivative of the logarithm 
of WIG which can in fact be viewed as the "finger print" of a dynamical critical phenomenon. 
\end{abstract}
\PACS{05.45.Tp, 89.65.Gh, 89.75.-k}
\noindent

\section{Introduction}\label{section:Introduct}
It seems that there are many distinct analogies between the dynamics and/or stochastics of complex physical and 
economical or even social systems \cite{RBAP,PB,MS,BP,KI,BR,DS,EPN,FS}. The methods and even algorithms that have 
been explored for description of physical phenomena become an effective background and inspiration for very 
productive methods used in analysis of economical data \cite{BK,GM}. 

In this paper we study an emerging market and more precisely, the historical Warsaw Stock Exchange (WSE) index WIG 
at a daily time horizon at the closing; we think that its dynamics is typical for an emerging financial market of 
small and moderate size. Our concept is to consider only well developed temporal (local) maxima of the 
{\it cusp}-like shape decorated with some oscillations (cf. peaks denoted by $A, a, B, C$ in 
Fig.\ref{figure:WIG1991-06_do_kwiec}) which cover the greater part of the whole time series. Our goal is to describe 
the slowing down of rising and relaxation processes within these temporal maxima by assuming a retarded feedback 
as a principal effect (except the rising path within the first local peak which, in principle, is exponential; 
cf. Fig.\ref{figure:WIGA_Dud}). This feedback is a reminiscence of investors' activity stimulated mainly by their 
observations of the empirical data in the past. 

\subsection{Inspiration}\label{section:Insp}

Our analysis was inspired by the non-Debye or 
non-exponential, fractional relaxation processes observed, for example, in stress-strain relaxation present in 
viscoelastic materials \cite{RH,KRS,GN}. The most commonly used empirical decay function for handling non-Debye 
relaxation processes in complex systems are described either by a Kohlraush-Williams-Watts (KWW) or a stretched 
exponential decay function \cite{RB} for short time-range $t \ll \tau $:
\begin{eqnarray}
f(t)\sim \exp\left(-\left(\frac{t}{\tau }\right)^{\alpha }\right),
\label{rown:KWW}
\end{eqnarray}
where $0<\alpha <1$, or by an asymptotic power-law (Nuttig law) for $t\gg \tau $:
\begin{eqnarray}
f(t)\sim \left(\frac{t}{\tau }\right)^{-\alpha }. 
\label{rown:Nl}
\end{eqnarray}
From expressions (\ref{rown:KWW}) and (\ref{rown:Nl}) we obtain for the change of $\ln f(t)$
per unit time\footnote{The logarithm of the index or price is the quantity playing a fundamental role 
in financial analysis both of stochastic or deterministic types, e.g. its time-derivative is the instantaneous 
interest rate or return per unit time.}
\begin{eqnarray}
\frac{d\ln f(t)}{dt}\sim 
\left\{
\begin{array}{cc}
\frac{1}{t^{1-\alpha }} & \mbox{for $t\ll \tau $} \\
\frac{1}{t} & \mbox{for $t\gg \tau $},
\end{array}
\right.
\label{rown:dfft}
\end{eqnarray}
which defines the power-law limits of the dynamics, i.e. it defines (for $t\ll \tau $) a universality class 
(characterized by dynamical, critical exponent $1-\alpha $) which, in fact, can be viewed as the "finger 
print" of a dynamical critical phenomenon. For such a peak the derivative diverges at 
the extremal point (i.e. at $t=0$) which justifies the name "sharp peak" used often in the literature in this 
context (cf. \cite{BR} and refs. therein). For a sufficiently wide empirical window it would be possible to observe 
a transition from the KWW to the Nutting behaviour and from one kind of power-law to another, correspondingly. 

Note that the non-exponential relaxation introduces memory, i.e. the underlying fundamental processes are of 
non-Markovian type. It was shown that a natural way to incorporate memory effects is fractional calculus. The 
power-law kernel defining the fractional relaxation equation involves a particularly long memory. The function 
which plays a dominating role in fractional relaxation problems is indeed the Mittag-Leffler (ML) function \cite{MK}  
\begin{eqnarray}
E_{\alpha }\left(-\left(\frac{t}{\tau }\right)^{\alpha }\right) = 
\sum_{n=0}^{\infty }\frac{(-(t/\tau )^{\alpha })^n}{\Gamma(1+\alpha n)},
\label{rown:eq1}
\end{eqnarray}
which is a natural generalisation of the exponential one. This function interpolates between cases 
(\ref{rown:KWW}) and (\ref{rown:Nl}) and plays a central role in our analysis. 

\section{The model}

Our phenomenological model of index dynamics consists of two stages:
\begin{itemize}
\item[(i)] Formulation of a linear ordinary differential equation of the first order with no feedback 
incorporated which describes evolution of an auxiliary, synthetic index only. 
\item[(ii)] A conjecture which transforms the above mentioned equation to a more general fractional form which 
already models the evolution of the empirical Warsaw Stock Exchange index. 
\end{itemize}
\begin{figure}[]
\begin{center}
\epsfig{file=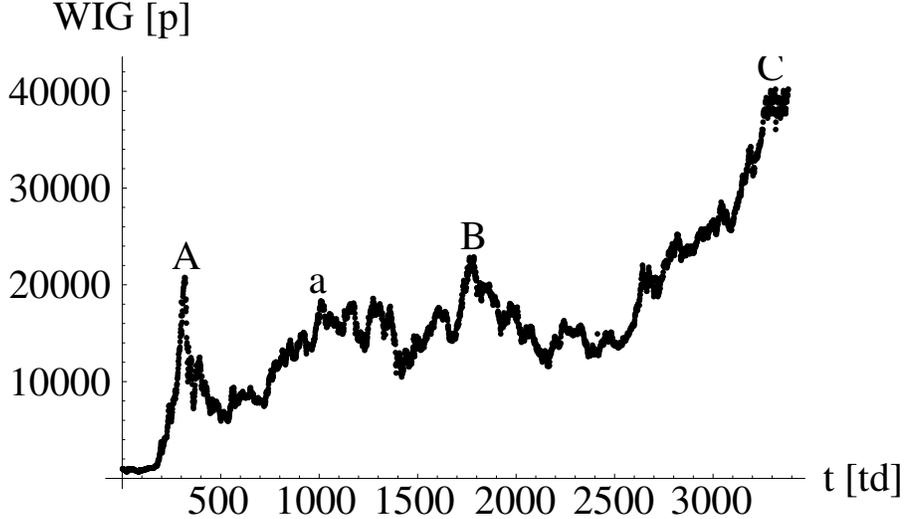, width=130mm}
\caption{The daily closing value of the Warsaw Stock Exchange index WIG (measured by conventional points (p)) 
from April 16, 1991 (the beginning of the Warsaw Stock Exchange activity) to the $30^{th}$ of March 2006; the 
presented time series consists of empirical data points for $3382$ trading days (td). The local maxima denoted 
as $A,\; a,\; B$ and $C$ are analysed further in the text.}
\label{figure:WIG1991-06_do_kwiec}
\end{center}
\end{figure}
The transition from (i) to (ii) means that the system is changed from an unrealistic to a realistic, complex 
one where the retarded feedback plays an essential role. By using this model we are able to describe the well 
developed temporal maxima present in daily time series defined by WIG activity (cf. four maxima $A,\; a,\; B$ 
and $C$ shown in Fig.\ref{figure:WIG1991-06_do_kwiec}).

\subsection{Evolution of WIG} 

The time-dependent value of WIG, $X(t)$, can be decomposed into two components which are recorded and can be 
even electronically accessible for traders: 
\begin{itemize}
\item[(a)] The instantaneous offset $U(t)\stackrel{\rm def.}{=}U_+(t)-U_-(t)$ between the temporal (total) demand 
$U_+(t)\ge 0$ for stocks defining WIG and their (total) temporal supply $U_-(t)\ge 0$. 
\item[(b)] The instantaneous volume trade $V(t)\stackrel{\rm def.}{=}min[U_+(t),U_-(t)]$ of stocks of the companies 
which consitute WIG.
\end{itemize}
Hence, we can write the following instantaneous superposition
\begin{eqnarray}
X(t)=A\cdot U(t)+B\cdot V(t),
\label{rown:XsV}
\end{eqnarray}
where $A(>0)$ and $B$ are referred to as coefficients of relocation. Note that paradoxally index 
$X(t)$\footnote{Note that $X(t)$ is defined here relative to some reference level so it can be, in general, 
both positive and negative.} can change even when the volume of trade, $V(t)$, vanishes, i.e. when the demand 
or supply vanishes. On the other hand, for vanishing $U(t)$ (when the demand is balanced by the supply) volume trade,
$V(t)$, can be still nonvanishing which leads to the change of WIG\footnote{Eq.(\ref{rown:XsV}) defines an additive 
variant of our model though a multiplicative one is also possible.}. 

To consider the dynamics (evolution) of WIG, we complete eq.(\ref{rown:XsV}) by the differential 
one which exhibits an instantaneous, therefore unrealistic, feedback to the financial market, namely 
\begin{eqnarray}
\frac{dV(t)}{dt}=C\cdot V(t)+D\cdot X(t)+E\cdot \frac{dX(t)}{dt},
\label{rown:dtUVX}
\end{eqnarray} 
where coefficients $C,\; D$ are rates while $E$ is again a kind of relocation coefficient. By combining  
eqs.(\ref{rown:XsV}) and (\ref{rown:dtUVX}) we eliminate the $V(t)$ variable and obtain (after integration), 
\begin{eqnarray}
X(t)-X(0)-sgn(C^{\prime })\; \tau _1^{-1}\;_0D_t^{-1}X(t)&=&-A^{\prime }\;sgn(C)\; \tau _0^{-1}\;_0D_t^{-1}U(t)
\nonumber \\
&+&A^{\prime }\; [U(t)-U(0)],  
\label{rown:cedt}
\end{eqnarray} 
where the definition of an inverse derivative of the first order was used; the definition of its general 
$n^{th}$ order version (for $n=1, 2, 3, \ldots $) has a useful form given by the Cauchy formula of repeated 
integration
\begin{eqnarray}
\;_0D_t^{-n}f(t)&\stackrel{\rm df.}{=}&\int_0^t dt_{n-1}\int_0^{t_{n-1}}dt_{n-2}\ldots \int_0^{t_1}f(t_0)dt_0
\nonumber \\
&=&\frac{1}{\Gamma (n)}\int_0^t dt^{\prime }(t-t^{\prime })^{n-1}f(t^{\prime }).
\label{rown:defD1}
\end{eqnarray} 
The combined coefficients
\begin{eqnarray}
A^{\prime }=\frac{A}{1-BE}, \; C^{\prime }=C\; \frac{1-B\; \frac{D}{C}}{1-BE}, 
\tau _0^{-1}\stackrel{\rm df.}{=}\; \mid C\mid ,\; \tau _1^{-1}\stackrel{\rm df.}{=}\; \mid C^{\prime }\mid ,
\label{rown:cc}
\end{eqnarray}
which are valid for $BE\not=1$; otherwise, instead of eq.(\ref{rown:cedt}) we would obtain a very special one. 

The integral eq.(\ref{rown:cedt}) defines the model which is an analog of the Zener one for viscoelastic solids 
\cite{NT} in which the stress ($U$) - strain ($X$) relationship\footnote{Usually the stress is denoted by $\sigma $ 
and strain by $\epsilon $.} is given originally by the linear first order differential equation \cite{GN}. 
Indeed, eq.(\ref{rown:cedt}) is the one which we generalize to the fractional form by applying its Maxwell 
formulation \cite{SB}. This formulation consists of a spring (obeying Hooke's law) and a dashpot (obeying Newton's 
law of viscosity) in series; this arrangement shows a simple spatial separation of the solid (elastic) and the 
fluid (viscous) aspects and it is too specific to describe real viscoelastic materials. However, the hierarchical
arrangement of a number (in general infinite) springs and dashpots is already sufficient \cite{SB}. Note, that
in our approach the spring defines a purely emotional or irrational investors' activity\footnote{In psychology more
often is used terminology 'affected driven activity' or 'authomatic activity'.} while the dashpot defines 
a purely rational one.  

\subsection{Conjecture}

There are several definitions of fractional differentiation and integration \cite{RH}. In what follows we are 
dealing strictly with the Liouville-Riemann (LR) fractional calculus. The fractional integration of arbitrary order 
$\alpha (>0)$ of function $f(t)$ is a straightforward generalisation of (\ref{rown:defD1}),
\begin{eqnarray}
\;_0D_t^{-\alpha }f(t)\stackrel{\rm df.}{=}\frac{1}{\Gamma (\alpha )}
\int_0^t dt^{\prime }\frac{f(t^{\prime })}{(t-t^{\prime })^{1-\alpha }},
\label{rown:fracdt}
\end{eqnarray}
where $\;_0D_t^{-\alpha }$ is the LR fractional integral operator of order $\alpha $ \cite{RH}. 

The fractional generalization of eq.(\ref{rown:cedt}) is performed by replacing expressions
$\tau _0^{-1}\; _0D_t^{-1}U(t)$ and $\tau _1^{-1}\; _0D_t^{-1}X(t)$ by  $\tau _0^{-\alpha }\; _0D_t^{-\alpha }U(t)$ 
and $\tau _1^{-\alpha }\; _0D_t^{-\alpha }X(t)$ ones, respectively, where the fractional (in general)
exponent $\alpha $ is a free but most important shape parameter. Hence, we obtain the fractional integral equation 
which is able to describe both independent paths (the rising and relaxation) of local temporary peaks of WIG:
\begin{eqnarray}
X(y)-X(0)&=&-\tau _1^{-\alpha }\;_0D_y^{-\alpha }X(y)
-A^{\prime }\;sgn(C)\; \tau _0^{-\alpha }\;_0D_y^{-\alpha }U(y) \nonumber \\ 
&+&A^{\prime }\; [U(y)-U(0)],  
\label{rown:fracedt}
\end{eqnarray} 
where it was tacitly assumed that $sgn(C^{\prime })=-1$, while the independent variable
\begin{eqnarray}
y=
\left\{
\begin{array}{cc}
t_{MAX}-t & \mbox{for rising: $t\le t_{MAX}$} \\
t-t_{MAX} & \mbox{for relaxation: $t\ge t_{MAX}$.}
\end{array}
\right.
\label{rown:yttMAXJ} 
\end{eqnarray}
As both paths of any peak are assumed as independent ones we consider all parameters present in 
eq.(\ref{rown:fracedt}) as (in general) different for different paths.

For relaxation the first term on the rhs of eq.(\ref{rown:fracedt}) describes feedback where the retarded value of 
index influences the present one; this value is sensitive here to the past one due to the algebraic, integral kernel. 
The second term on the rhs of eq.(\ref{rown:fracedt}) desribes explicitly a financial market retarded influence on 
the index (or the stock price); the third term gives the instantaneous influence. However, for the rising path the
situation is more complicated and eq.(\ref{rown:fracedt}) constitutes only a formal, convenient way to describe it. 
As we prove, the first and third terms constitute mainly the basis of a dynamical structure of the local (temporal) 
maximum of WIG.

\subsection{Relaxation and rising fractional differential equations}

To make a step towards the interpretation, we define the fractional differentiation of order $\gamma (>0)$ 
\begin{eqnarray}
\;_0D_t^{\gamma }f(t)\stackrel{\rm df.}{=}\frac{d^n}{dt^n}\left(\;_0D_t^{\gamma -n}f(t)\right),
\label{rown:fracbdt}
\end{eqnarray}
which is considered to be composed of a fractional integration of the order 
$\alpha \stackrel{\rm def.}{=}n-\gamma \; (-1\le \gamma -n<0)$ followed by an ordinary differentiation of order $n$.
Now, by ordinary differentiation of the first order of eq.(\ref{rown:fracedt}) and 
by applying definition (\ref{rown:fracbdt}) we obtain the linear inhomogeneous fractional differential 
equation
\begin{eqnarray}
\frac{dX(y)}{dy}&=&-\tau _1^{-\alpha }\;_0D_y^{1-\alpha }
X(y)-A^{\prime }\;sgn(C)\; \tau _0^{-\alpha }\;_0D_y^{1-\alpha }U(y) \nonumber \\
&+&A^{\prime }\; \frac{dU(y)}{dy}, 
\label{rown:dfracedt}
\end{eqnarray} 
which describes well both paths of the studied peaks. 

\subsubsection{Free fractional relaxation: the reference case} 

We found that the well developed local maxima of the index can be fitted (except for the left-hand side of the first 
one and up to their oscillations and fluctuations) by an intermediate part of the ML function. In our case we 
obtained several values of exponent $\alpha $ for WIG's maxima and almost all of them (except one) are smaller than 
$0.5$; note that the left-hand side of the first maximum is well fitted by the usual exponential function (or by
assuming $\alpha =1$ in the MF function).   

In other words, the relaxation of almost all WIG local maxima can be described by the fractional relaxation equation
by setting in eq.(\ref{rown:dfracedt}) coefficient $A^{\prime }=0$. Such a simplified equation is, of course, 
a fractional generalization of the standard relaxation equation whose solution has indeed
the form (\ref{rown:eq1}) (where variable $t$ is replaced by $y$ and parameter $\tau $ by corresponding
$\tau _1$ one). This solution is considered here only as a reference case. 

\subsection{Full Solution of the Fractional Initial Value Problem}

We solve the fractional initial value problem (\ref{rown:dfracedt}) by assuming that
\begin{eqnarray}
U(y)=\frac{U(0)}{2}\; [\exp(\imath \; (\omega -\Delta \omega )\; y) + 
\exp(\imath \; (\omega +\Delta \omega )\; y)], 
\label{rown:expat}
\end{eqnarray}
(where parameters $U(0), \omega, \Delta \omega >0$) and by applying the Laplace transform of a fractional integral 
operator. Namely, the Laplace transformation of eq.(\ref{rown:fracedt}) yields 
\begin{eqnarray}
\tilde{X}(s)=A^{\prime }\; \frac{1-sgn(C)\tau _0^{-\alpha }s^{-\alpha }}{1+\tau _1^{-\alpha }s^{-\alpha }}
\; {\tilde{U}(s)}+[X(0)-A^{\prime }\; U(0)]\; \frac{s^{-1}}{1+\tau _1^{-\alpha }s^{-\alpha }},  
\label{rown:transLapX}
\end{eqnarray}
where the Laplace transform of the LR fractional integral operator 
was applied here. By introducing the Laplace transform of (\ref{rown:expat})  
into eq.(\ref{rown:transLapX}) and by using the inverse Laplace transformation in the time domain we can obtain 
the real part of the solution.
However, to compare the prediction of our model with empirical data it is sufficient to use only the lowest order 
terms in the exact solution, i.e. it is suficient to use the following approximate solution
\begin{eqnarray}
\Re X(y)&\approx &\left[X(0)+A^{\prime }\; U(0)\; sgn(C)
\left(\frac{\tau _0}{\tau _1}\right)^{-\alpha }\right]
E_{\alpha }\left(-\left(\frac{y}{\tau _1}\right)^{\alpha }\right)+ \nonumber \\
&-&A^{\prime }\; U(0)\; sgn(C)
\left(\frac{\tau _0}{\tau _1}\right)^{-\alpha }\cos(\omega \; y)\cos(\Delta \omega \; y),
\label{rown:approxfde}
\end{eqnarray}
since the parameters $\omega ,\; \Delta \omega $, which additionaly multiply the integral terms, were found  
to be so small that the integral terms are negligible.

\subsection{Comparison with empirical data and discussion}

\begin{figure}[] 
\begin{center}
\epsfig{file=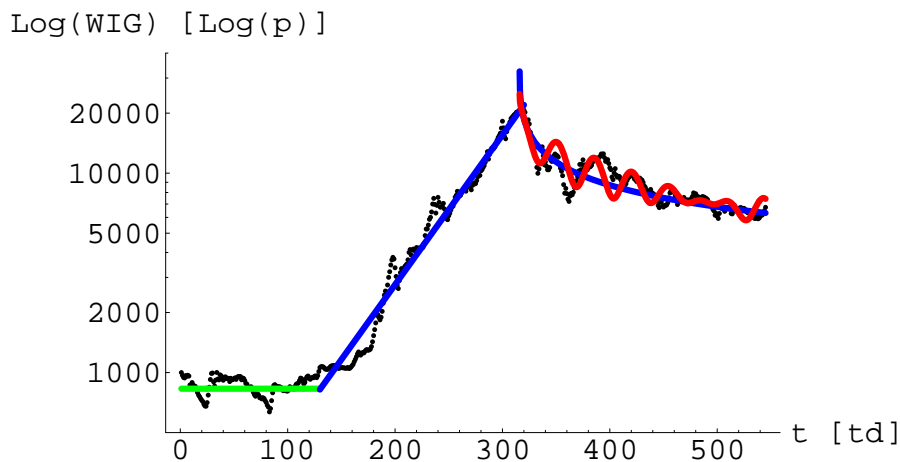, width=120mm}
\caption{The temporal, local maximum $A$ (cf. Fig.\ref{figure:WIG1991-06_do_kwiec}): the evolution of the daily 
closing price of WIG in the semi-logarithmic scale (dots) from the beginning of the WSE activity to the end of 
the first maximum on March 30, 1995; the presented time series consists of data points for $545$ trading days. 
The tangent solid line was fitted to the left-hand slope of the empirical maximum. The curve fitted to the 
right-hand slope was obtained by using eq.(\ref{rown:approxfde}). The values of the corresponding key parameters are 
presented in Table \ref{table:tab1}.}
\label{figure:WIGA_Dud}
\end{center}
\end{figure}

\begin{figure}[]
\begin{center}
\epsfig{file=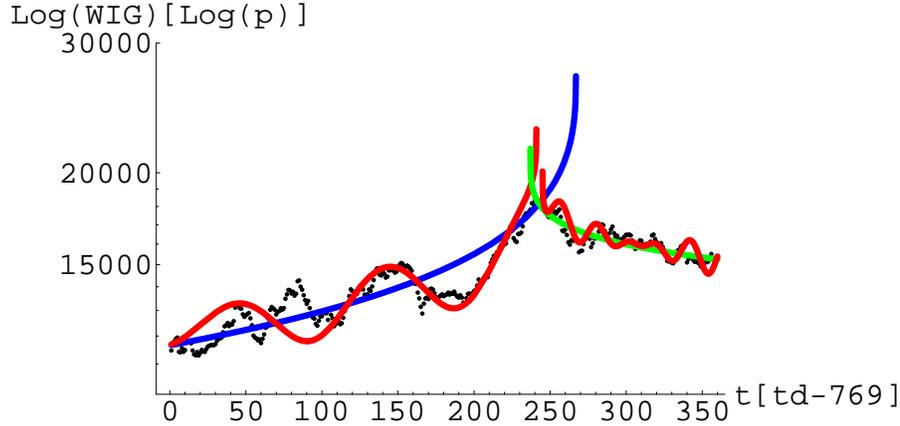, width=120mm}
\caption{The local maximum $a$ (cf. Fig.\ref{figure:WIG1991-06_do_kwiec}): the evolution of the daily closing value 
of WIG in the semi-logarithmic scale (dots) dated from the $770^{th}$ trading day (the beginning of this local 
maximum) to the $1130^{th}$ assumed as the end of this maximum range; the time series consists of 
data points for $360$ trading days. The curves fitted to left-hand and right-hand paths of the maximum were obtained 
by using eq.(\ref{rown:approxfde}); the corresponding key parameters are presented in Table \ref{table:tab3}.} 
\label{figure:WIGCent_Dud_17_01_06}
\end{center}
\end{figure}

\begin{figure}[]
\begin{center}
\epsfig{file=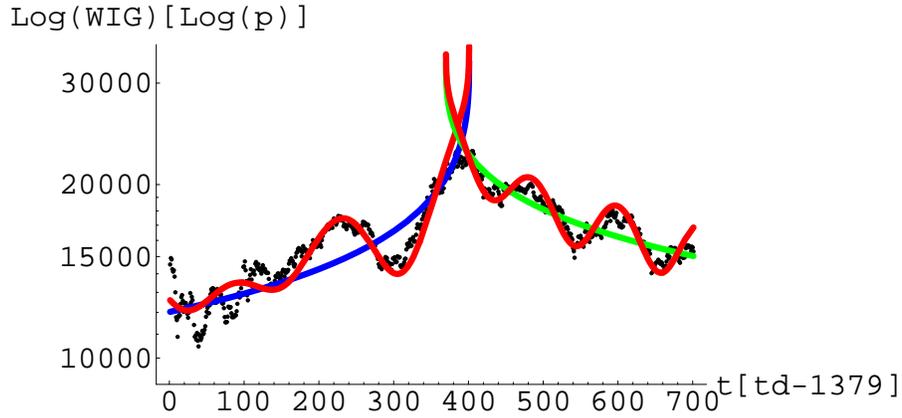, width=120mm}
\caption{The local maximum $B$ (cf. Fig.\ref{figure:WIG1991-06_do_kwiec}): the evolution of the daily closing value 
of WIG in the semi-logarithmic scale (dots) dated from the $1380^{th}$ trading day (the beginning of the local 
maximum) to the $2080^{th}$ day assumed as the end of this maximum range; the presented time series consists of 
data points for $700$ trading days. The curves fitted to both slopes of the empirical maximum were obtained by using 
eq.(\ref{rown:approxfde}). The corresponding key parameters are presented in Table \ref{table:tab5}.} 
\label{figure:WIGB_Dud_17_01_06}
\end{center}
\end{figure}

\begin{figure}[]
\begin{center}
\epsfig{file=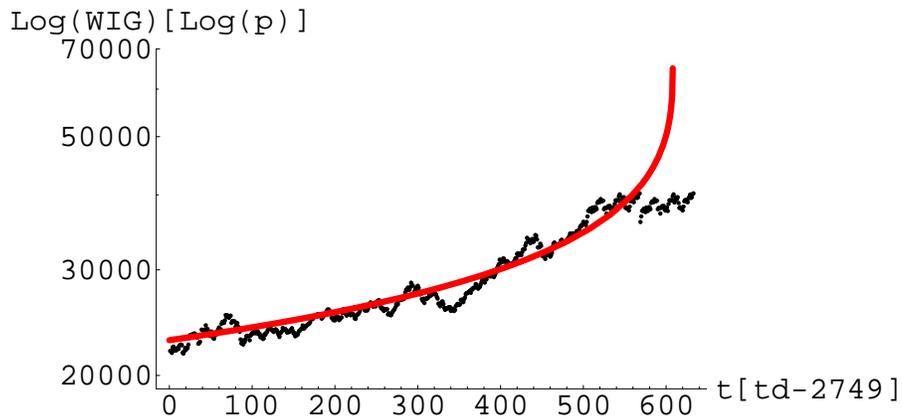, width=120mm}
\caption{The local maximum $C$ (cf. Fig.\ref{figure:WIG1991-06_do_kwiec}): the evolution of the daily closing value 
of WIG (dots) dated from the $2750^{th}$ trading day (the beginning of the local 
maximum) to the $3382^{nd}$ day assumed as the end of the maximum range; the presented time series consists of 
data points for $633$ trading days. The curve fitted to the left-hand slope of the empirical maximum was obtained 
by using eq.(\ref{rown:approxfde}). The corresponding key parameters are presented in Table \ref{table:tab7}.} 
\label{figure:WIGD_06_kw}
\end{center}
\end{figure}

In Figs.\ref{figure:WIGA_Dud}-\ref{figure:WIGD_06_kw} we compared the empirical data defining WIG's local maxima 
(denoted as $A,\; a,\; B,$ and $C$ in Fig.\ref{figure:WIG1991-06_do_kwiec}) with the predictions given by formula 
(\ref{rown:approxfde}). The monotonic curves (obtained by using only the first term) present free solutions while the
oscillating curves (obtained by using the whole expression) the full ones, i.e. the free solutions decorated with 
mono-frequency oscillations (rising and falling paths of peaks $a$ and $B$, respectively) or wiggles (right-hand paths 
of peaks $A$ and $a$ as well as left-hand path of peak $B$). The values of the key parameters which we obtained are 
shown in Tables \ref{table:tab1}-\ref{table:tab7}\footnote{Note that numbers shown without errors mean that errors 
are negligibly small.}. 
\begin{table}
\begin{center}
\caption{First set of parameters describing the temporal maximum $A$}
\begin{tabular}{|c||c|c|c|c|c|}
\hline
Process & $\alpha $ & $\tau _1\; [td]$ & $t_{MAX}\; [td]$ & $\omega \; [td^{-1}]$ & $\Delta \omega \; [td^{-1}]$ \\
\hline
\hline
Rising & $1$ & $58\pm 1$ & $320\pm 1$ & $-$ & $-$ \\
\hline
Relaxation & $0.41$ & $176\pm 1$ & $316\pm 1$ & $0.18$ & $0.009$ \\
\hline
\end{tabular}
\label{table:tab1}
\end{center}
\end{table}

\begin{table}
\begin{center}
\caption{First set of parameters describing the temporal maximum $a$}
\begin{tabular}{|c||c|c|c|c|c|}
\hline
Process & $\alpha $ & $\tau _1\; [td]$ & $t_{MAX}\; [td]$ & $\omega \; [td^{-1}]$ & $\Delta \omega \; [td^{-1}]$ \\
\hline
\hline
Rising & $0.31$ & $721.5\pm 1$ & $241\pm 1$ & $0.063$ & $0.005$ \\
\hline
Relaxation & $0.25$ & $6830\pm 10$ & $245\pm 1$ & $0.26$ & $0.025$ \\
\hline
\end{tabular}
\label{table:tab3}
\end{center}
\end{table}

\begin{table}
\begin{center}
\caption{First set of parameters describing the temporal maximum $B$}
\begin{tabular}{|c||c|c|c|c|c|}
\hline
Process & $\alpha $ & $\tau _1\; [td]$ & $t_{MAX}\; [td]$ & $\omega \; [td^{-1}]$ & $\Delta \omega \; [td^{-1}]$ \\
\hline
\hline
Rising & $0.40$ & $164\pm 1$ & $401\pm 1$ & $0.035$ & $0.005$ \\
\hline
Relaxation & $0.39$ & $366.5\pm 1$ & $370\pm 1$ & $0.055$ & $-$ \\
\hline
\end{tabular}
\label{table:tab5}
\end{center}
\end{table}

\begin{table}
\begin{center}
\caption{First set of parameters describing the temporal maximum $C$}
\begin{tabular}{|c||c|c|c|}
\hline
Process & $\alpha $ & $\tau _1\; [td]$ & $t_{MAX}\; [td]$ \\
\hline
\hline
Rising & $0.40$ & $249.2\pm 1$ & $608\pm 1$ \\
\hline
\end{tabular}
\label{table:tab7}
\end{center}
\end{table}

\begin{figure}[]
\begin{center}
\epsfig{file=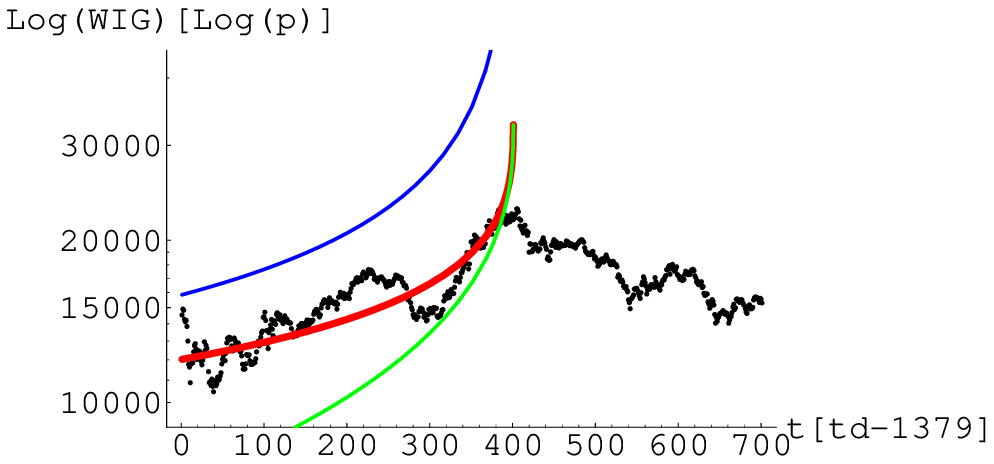, width=120mm}
\caption{The local WIG's maximum $B$ (cf. Fig.\ref{figure:WIG1991-06_do_kwiec}): the analysis of the evolution of the 
daily closing value of the left-hand path presented in the semi-logarithmic scale (dots). The middle 
solid curve fitted to the left-hand slope of the empirical maximum was taken from Fig.\ref{figure:WIGB_Dud_17_01_06}, 
while the lower and upper curves are the corresponding limits for the short time given by the streched exponential- 
or the KWW law (\ref{rown:KWW}), and for asymptotic time the power- or the Nutting law is given by 
(\ref{rown:Nl}).}
\label{figure:WIGBLAnal_New}
\end{center}
\end{figure}

In Fig.\ref{figure:WIGBLAnal_New} are presented: (i) the Mittag-Leffler function fitted, for example, to
the left-hand path of the empirical maximum $B$ (this is the free solution taken, in fact, from 
Fig.\ref{figure:WIGB_Dud_17_01_06}), and corresponding (ii) KWW law (lower curve) as well as (iii) the Nutting 
law (upper curve). This is a typical situation valid both for the left- and right-hand paths of all considered peaks. 
It is characteristic that none of the peaks reached the fully developed scaling region of the return per unit time 
$\frac{d\ln X(y)}{dy}$ (cf. section \ref{section:Insp}). In this sense the considered peaks have a precritical 
character.   

There are several other features common for all peaks which should be noted: 
\begin{itemize}
\item[(i)] Both paths of any peak can be considered as independent ones and the location of turning point (extremum)  
as a random one.
\item[(ii)] The considered peaks are asymmetric since:
\begin{itemize}
\item[$\bullet $] the exponent $\alpha $, which characterizes the left-hand paths of any peak, is larger than the 
analogous one characterizing the right-hand one,
\item[$\bullet $] parameter $\tau _1$ (defining the time unit) of the left-hand path is smaller than the 
corresponding one for the right-hand path. 
\end{itemize}
\item[(iii)] The location of the extremal point $t_{MAX}$ given by expression (\ref{rown:approxfde}) is (in general) 
different for left- and right-hand paths of each peak.
\end{itemize}
Moreover, a frequency modulation signal is necessary and/or the influence of the signal outside the maximum
should be taken into account to describe the beginning of the left-hand paths of peaks $a$ and 
$B$ (cf. Figs.\ref{figure:WIGCent_Dud_17_01_06} and \ref{figure:WIGB_Dud_17_01_06}).

Concluding, we suggest that our approach can rationally decrease the risk of investment on the stock market 
since it is able to warn the investors before the stock market reaches a critical region. \\
\\
We thank prof. Piotr Jaworski from the Institute of Mathematics at the Warsaw University for his helpful discussion.

\end{document}